\documentclass[aps,prl,twocolumn,superscriptaddress,showpacs]{revtex4}

\usepackage{graphicx}
\usepackage{dcolumn}
\usepackage{bm}
\usepackage{color}

\begin{document}

\title{Self-energy effects and electron-phonon coupling in Fe-As superconductors}

\author{K.-Y. Choi$^{1}$, P. Lemmens$^2$, I. Eremin$^{3,4}$, G. Zwicknagl$^{4}$, H. Berger$^5$, G. L. Sun$^{6}$, D. L. Sun$^{6}$, and C. T. Lin$^{6}$}
\address{Department of Physics, Chung-Ang University, Dongjak-Gu, Seoul 156-756, Republic of Korea}
\address{Institute for Condensed Matter Physics, TU Braunschweig, D-38106 Braunschweig, Germany}
\address{Max-Planck-Institut f\"ur Physik komplexer Systeme, D-01187 Dresden, Germany}
\address{Institut f\"ur Mathematische Physik, TU Braunschweig, D-38106 Braunschweig, Germany}
\address{Institute de Physique de la Matiere Complexe, EPFL, CH-1015 Lausanne, Switzerland}
\address{Max-Planck-Institut f\"ur Festk\"orperforschung, Heisenbergstr. 1,
D-70569 Stuttgart, Germany}

\date{\today}
\begin{abstract}
We report on Raman scattering experiments of the undoped SrFe$_{2}$As$_{2}$ and
superconducting Sr$_{0.85}$K$_{0.15}$Fe$_{2}$As$_{2}$ ($T_c=28~K$) and Ba$_{0.72}$K$_{0.28}$Fe$_{2}$As$_{2}$ ($T_c=32~K$) single crystals. The frequency and linewidth of the B$_{1g}$ mode at 210 cm$^{-1}$ exhibits an appreciable temperature dependence induced by the superconducting and spin density wave transitions. We give estimates of the electron-phonon coupling related to this renormalization. In addition, we observe a pronounced quasi-elastic Raman response for the undoped compound, suggesting persisting magnetic fluctuations to low temperatures.
In the superconducting state the renormalization of an electronic continuum is observed with a threshold energy of $61~\mbox{cm}^{-1}$.

\end{abstract}


\maketitle



The recent discovery of superconductivity in iron arsenide compounds containing Fe$_2$As$_2$ layers with $T_c$ higher than 50~K has stimulated enormous research activities~\cite{Kamihara,Takahashi,Chen,Ren,Rotter}. This is largely owed to the tantalizing possibility of finding a unified framework and description of a diverse range of unconventional superconducting materials.

Similarly to the high-$T_c$ cuprates there is evidence for the importance of many-body effects.
Inelastic neutron scattering (INS) measurements have uncovered a magnetic resonance peak in superconducting Ba$_{0.6}$K$_{0.4}$Fe$_{2}$As$_{2}$~\cite{Christianson}. The electron-phonon coupling is estimated to be a factor 5 too weak to reproduce the observed transition temperature, even taking account multiband effects~\cite{Boeri}. Besides, the spin-density-wave (SDW) and the phase diagram with doping indicate an intimate relation between magnetism and superconductivity.

Combining the specific band structure of the Fe pnictides~\cite{Mazin} with antiferromagnetic correlations is proposed to lead to a $s^{\pm}$-wave pairing state ($\Delta_{\bf k} = \frac{\Delta_0}{2}\left(\cos k_x +\cos k_y \right)$). The $s^{\pm}$-gap symmetry assumes the change of sign of the superconducting gap between electron and hole pockets and the gap's magnitudes differ slightly on the hole and electron pockets. There is also a fictitious line of nodes between pockets which, however, does not cross the Fermi surface.  This scenario is further supported by several theoretical works  ~\cite{Kuroki,Bang,Korshunov}. However, the experimental situation is still less evident.
ARPES, INS, and Andreev spectroscopy measurements are consistent with $s^{\pm}$-wave symmetry~\cite{Chen,Christianson,Ding}, but there exist some other like power-law temperature dependencies of the penetration depth \cite{prozorov,carrington} which cannot be easily fitted by the nodeless $s^{\pm}$-wave symmetry \cite{vorontsov}. To resolve this peculiarity several mechanisms have been proposed. All of them are based on the assumption that due to relatively large intraband Coulomb repulsion within each of the pockets the superconducting gap forms a node within at least one of them. Depending on the details of the model it can be obtained either by inclusion of the higher harmonics into extended $s$-wave gap with $\cos k_x/2 \cos k_y/2$ dependence which results in the nodal lines crossing electron pockets centered around the $M$-point \cite{chubukov_nod,meyer} or
by assuming different symmetries like the $d_{xy}$-wave ($\sin k_x \sin k_y$) or $d_{x^2-y^2}$-wave ($\cos k_x - \cos k_y$) symmetries \cite{moreo,aoki}.

The prevailing trust in a magnetically mediated pairing mechanism with moderate electron-phonon coupling
has been further confirmed by the observation of an inverse isotope coefficient on iron $\alpha\sim -0.18$ of Ba$_{1-x}$K$_{x}$Fe$_{2}$As$_{2}$  compound\cite{Liu,Bang08}. At the same time, Eschrig~\cite{Eschrig} asserted that an in-plane B$_{1g}$  mode might have a large electron-phonon coupling constant, which is more relevant to superconductivity than the averaged modes.

Raman spectroscopy is an experimental tool of choice for studying the superconducting gap symmetry as well as the strength of electron-phonon coupling \cite{cobaltates}. In fact, earlier Raman works have investigated the SDW and superconducting properties of pnictides~\cite{Litvinchuk,Choi,Granath,Rahlenbeck}. Although SDW and superconductivity related phonon anomalies are observed, neither electron-phonon coupling nor a superconductivity-induced renormalization of an electronic Raman continuum have been discussed in a combined experimental/theoretical study.

In this paper, we provide spectroscopic information on the B$_{1g}$ phonon renormalization as well as on the renormalization of an electronic continuum upon entering into the superconducting phase. In particular, we address
the electron-phonon coupling constant and the superconducting gap.

For Raman scattering experiments, single crystals with dimensions of $3\times 2\times0.2~\mbox{mm}^3$  were used. The samples were grown by the flux growth method using solvents in zirconia crucibles under argon atmosphere~\cite{Sun}. The superconducting transition temperatures of $T_c=28~K$ and $T_c=32~K$ for Sr$_{0.85}$K$_{0.15}$Fe$_{2}$As$_{2}$ (SKFA) and Ba$_{0.72}$K$_{0.28}$Fe$_{2}$As$_{2}$ (BKFA), respectively, were identified by transport and resistivity measurements. As a lattice reference we used the diamagnetic, isostructural compound FeAs$_2$. The samples were kept in the vacuum of an optical cryostat equipped with a closed cycle refrigerator. Raman scattering experiments were performed using the excitation line $\lambda= 532$~nm (Nd:YAG solid-state Laser) in a quasi-backscattering geometry.
A laser power of 5~mW was focused to a 0.1~mm diameter spot on the surface of the single crystal. The scattered spectra were collected by a DILOR-XY triple spectrometer and a nitrogen cooled charge-coupled device detector.

\begin{figure}[tbp]
\linespread{1}
\par
\includegraphics[width=8cm]{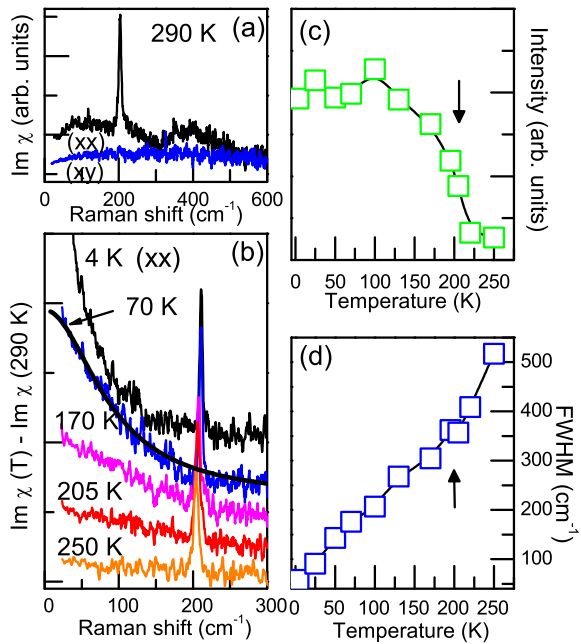}
\par
\caption{(a) Raman response $\mbox{Im}\chi$ of SrFe$_{2}$As$_{2}$ for (xx) and (xy) polarizations
at 290~K. (b) Temperature dependence of Raman response obtained by subtracting $\mbox{Im}\chi (K)$
from $\mbox{Im}\chi (290K)$. The solid line is a fit to a Lorentzian profile.
(c) Temperature dependence of the scattering intensity.
(d) Temperature dependence of the full width at half maximum.} \label{fig:1}
\end{figure}
Figure~1(a) shows the Raman response $\mbox{Im}\chi$ of the undoped SrFe$_{2}$As$_{2}$
for (xx) and (xy) polarizations at 290~K, which is corrected by the Bose thermal factor $[1+n(\omega)]
=[1-\exp(-\hbar\omega/k_BT)]^{-1}$ from the measured Raman scattering intensity. At room temperature we observe a single peak at $203~\mbox{cm}^{-1}$ in the (xx) polarization. This is part of four symmetry-allowed modes; $\Gamma_{Raman}=A_{1g}(x^2+y^2,z^2) + B_{1g}(x^2-y^2)+ 2E_{g}(xz,yz)$~\cite{Litvinchuk,Choi}.  The $203~\mbox{cm}^{-1}$ mode is assigned to a B$_{1g}$  mode and  corresponds to a displacement of Fe atoms along the $c$ axis. The room-temperature Raman response exhibits a structured background, which might be composed of a phonon density of states and weak electronic Raman scattering. To study the intrinsic electronic Raman response $\mbox{Im}\chi (T=290K)$ is subtracted from $\mbox{Im}\chi (T)$. The resulting Raman response is plotted in Fig.~1(b) as a function
of temperature.

We observe an elastic-scattering maximum, which is well described by a Lorentzian profile, $\mbox{Im}\chi
\propto A\Gamma/(\omega^2+\Gamma^2)$, where A is the scattering intensity and $\Gamma$ is the full width at half maximum. With decreasing temperature the scattering intensity grows steeply through $T_s$ of the SDW and then shows more or less saturation for temperature below 130 K [see Fig.~1(c)]. The full width at half maximum tends to go to zero quasilinearly upon cooling [see Fig.~1(d)]. The central maximum can arise from the decay of a soft mode into acoustic modes or phonon density fluctuations in the presence of the structural phase transition. In our case, however, this does not give a dominant contribution because the structural phase transition is of first order. Actually, the intensity of the elastic maximum does not diverge at $T_s$. Instead, it resembles to the temperature dependence of
the elastic neutron scattering intensity at the AF superlattice reflection~\cite{Zhao}. Thus, it is ascribed to light scattering by low energy magnetic excitations. The quasilinear dependence of the linewidth suggests strong magnetic fluctuations at low energies and thus a proximity to a quantum phase transition~\cite{Uhrig,QCPreview}.

\begin{figure}[tbp]
\linespread{1}
\par
\includegraphics[width=7cm]{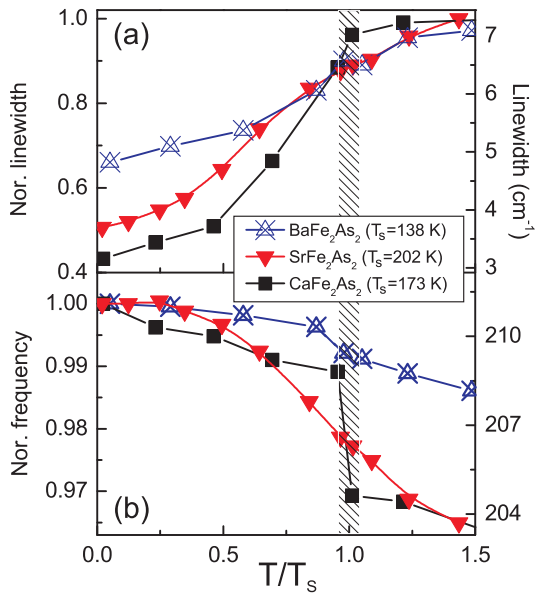}
\par
\caption{(a) Temperature dependence of the normalized linewidth and (b) peak position of the $203~\mbox{cm}^{-1}$ mode on a reduced temperature scale, $T/T_s$. For comparison, SrFe$_{2}$As$_{2}$ ($T_s=202~K$) is presented together with BaFe$_{2}$As$_{2}$ ($T_s=138~K$) and CaFe$_{2}$As$_{2}$ ($T_s=173~K$). The dashed, dotted, and solid lines are guides to the eye.} \label{fig:2}
\end{figure}

In order to understand the evolution of the SDW state the temperature dependence of the $203~\mbox{cm}^{-1}$ B$_{1g}$ mode was analyzed by a Lorentzian profile.  In Fig.~2 the results are summarized together with BaFe$_{2}$As$_{2}$ ($T_s=138~K$) and CaFe$_{2}$As$_{2}$ ($T_s=173~K$). We note that this mode is susceptible to any change of the Fe-$d$ states around the Fermi level.

The phonon frequency and linewidth show characteristic anomalies in the temperature dependence. In all samples we observe a jump of the phonon frequency and a narrowing of the linewidth below $T_s$. The discontinuous change of the frequency and lifetime at $T_s$ is largely rounded off for AFe$_2$As$_2$ (A=Sr,Ba) due to impurities~\cite{Ni08}. Similar phonon anomalies are observed in the A$_{1g}$ mode~\cite{Rahlenbeck}. Thus, the narrowing of the linewidth below $T_s$ should be attributed to the formation of the SDW state~\cite{tellurite}. Actually, the loss of the density of states on the Fermi surface can explain a longer phonon lifetime. However, we find no evidence for a depletion of the electronic Raman scattering. This suggests that the SDW gap is not fully opened on the Fermi surface.

\begin{figure}[tbp]
\linespread{1}
\par
\includegraphics[width=7.5cm]{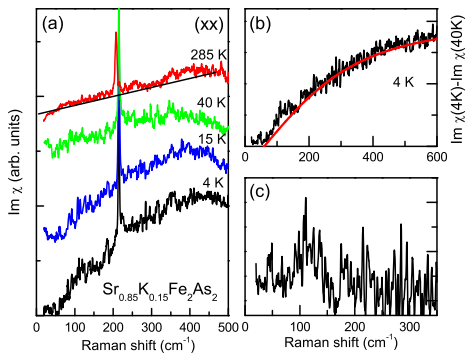}
\par
\caption{(a) The Raman response spectra of the SKFA ($T_c=28~K$) single crystal in the (xx) polarization at 285, 40, 15, and 4 K. The spectra are systematically shifted vertically. (b) The electronic Raman response obtained by subtracting
the 40~K spectrum from the 4 K one. The solid line is a fit by $\omega/\sqrt{a+b\omega^2}H(\omega-2\Delta)$ where $H(n)$ is the Heaviside step function with a gap cutoff $2\Delta=61 ~\mbox{cm}^{-1}$.
(c) Subtraction of the $Im~\chi(4K)- Im~\chi(40K)$ spectrum by the fitted curve.} \label{fig:3}
\end{figure}

In Fig.~3(a) we show the electronic Raman response $Im\chi$ of the superconducting Sr$_{0.85}$K$_{0.15}$Fe$_{2}$As$_{2}$ in (xx) polarization. At room temperature we observe a weakly linear-dependent electronic background, which is typical for a paramagnetic metallic state. Upon cooling through $T_c$ a superconductivity-induced renormalization of the electronic continuum shows up at low energies. However, no pronounced pair-breaking peak is visible. This might be due to surface impurities and disorders, which lead to Rayleigh scattering and luminescence. To minimize the extrinsic effects, we subtract the normal state spectrum at 40~K from the superconducting state one at 4~K. The resulting spectrum is well fitted by $\omega/\sqrt{a+b\omega^2}H(\omega-2\Delta)$ where $H(n)$ is the Heaviside step function with a gap cutoff $2\Delta=61~\mbox{cm}^{-1}$ [See Fig.~3(b)].

We note that for an isotropic $s$-wave gap the Raman scattering intensity is totally suppressed below a pair breaking peak of $2\Delta$. In contrast, an unconventional superconductor with nodes in the gap function has a substantial scattering intensity for energies below the gap. In order to identify superconductivity-related excitations from the $Im~\chi(4K)- Im~\chi(40K)$ spectrum the fitted curve is subtracted. The result is plotted in Fig.~3(c). Although contaminated by strong noise, the Raman residual response appears to be almost negligible below a threshold energy of $61~\mbox{cm}^{-1}$.  This feature seems to be compatible with a nodeless gap function. In addition, we observe a weak, broad feature at about 120~cm$^{-1}$. This would correspond to $2\Delta/k_BT_c\sim 5.3$. Most of the superconducting gap studies on the hole-doped AFe$_2$As$_2$ family evidence multiple nodeless gaps. Our data fall into the reported gap range with  $2\Delta/k_BT_c\sim 6-8$~\cite{Samuely}. For a $s^{+}$-wave gap, a theoretical work \cite{Chubukov_raman} proposed that the A$_{1g}$ Raman intensity has a resonance peak below $2\Delta$. The (xx) scattering geometry contains an A$_{1g}$ symmetry but there is no apparent hint for a resonance peak. Since the superconducting coherence peak itself is very weak, it is indispensable to study a cleaner sample to arrive at a definite conclusion on the superconducting gap symmetry.

\begin{figure}[tbp]
\linespread{1}
\par
\includegraphics[width=7.5cm]{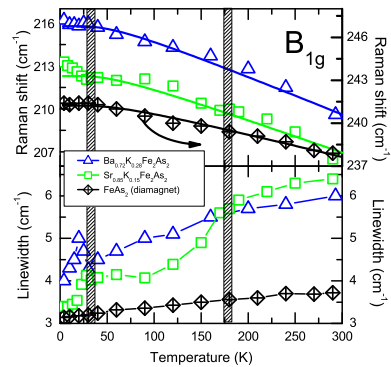}
\par
\caption{Temperature dependence of the peak position (Upper panel) and linewidth (Lower panel) of the B$_{1g}$ mode for SKFA ($T_c=28~K$) and BKFA ($T_c=32~K$) single crystals. For comparison, results of the diamagnetic, isostructural compound FeAs$_2$ are added. The solid lines are a fit to anharmonic phonon decay process. The striped bar highlights the superconducting and spin density wave transition regimes, respectively.} \label{fig:4}
\end{figure}

Shown in Fig.~4 is the temperature dependence of the frequency and linewidth of the B$_{1g}$ mode for superconducting SKFA and BKFA.  In order to identify possible electron-phonon contributions to the frequency shift and line broadening, data of the diamagnetic isostructural compound FeAs$_2$ is presented as well. Its temperature dependence of the phonon frequency is well given by simple phonon-phonon decay processes~\cite{Balkanski},
$$\omega_{ph}(T)=\omega_0+C[1+2/\exp(-\hbar\omega_0/2k_BT)-1]^{-1}$$ where $\omega_0$ is the zero-temperature frequency of the B$_{1g}$ mode and $C$ describes anharmonic phonon decay processes. BKFA follows largely the modeled anharmonic behavior. For SKFA with a higher SDW transition temperature, however, we can find sizable deviations from the anharmonic profile. The frequency undergoes a small hardening by 1~cm$^{-1}$ at $T_c$. This means that although the overall temperature dependence of the frequency is determined by the lattice anharmonicity, there are nonnegligible superconductivity-induced self-energy contributions.

To analyze the effect of superconductivity on the renormalization of the B$_{1g}$ phonons we employ the
four-band model proposed previously\cite{Korshunov} for the folded Brillouin Zone. The unit cell contains two Fe and two As atoms and the band structure predicts a Fermi surface consisting of two hole ($\alpha$) pockets centered around the $\Gamma-$point and two electron ($\beta$) pockets centered around the $M$-point, respectively.

Without taking into account vertex corrections, the renormalization of the optical phonons is determined by the Dyson equation:
\begin{equation}
D^{-1}({\bf q},\omega)=D^{-1}_0(\omega)- g_{\bf q}^{2} \Pi({\bf q},\omega),
\label{eq:dyson}
\end{equation}
where $D_0 (\omega) = \frac{2\omega_\Gamma}{\omega^2-\omega_\Gamma^{2} +
i\delta}$ is the momentum-independent bare phonon propagator and $g_{\bf q}$ is the corresponding electron-phonon coupling constant. The polarization
operator is given by
\begin{equation}
\Pi({\bf q}, \omega) = - i\int \mbox{Tr} \left[\tau_3 G ({\bf k+q}, \Omega + \omega)\tau_3 G ({\bf k}, \Omega)\right] \frac{d^2 k d\Omega}{(2\pi)^3}, \label{eq:Pi}
\end{equation}
where $G({\bf k},\omega)=\frac{\omega I +\varepsilon_{\bf k} \tau_3 +\Delta_{\bf k} \tau_1}{\omega^2-E_{\bf k}^2+i\delta}$ is the propagator, $E_{\bf k}^2=\varepsilon^2_{\bf k}+\Delta_{\bf k}^2$ is the energy dispersion in the superconducting state, $\varepsilon_{\bf k}$ are the tight-binding energies for the $\alpha$ and $\beta-$bands  \cite{Korshunov}, and $\Delta_{\bf k}$ is a (momentum-dependent) superconducting gap.

\begin{figure}[tbp]
\linespread{1}
\par
\includegraphics[width=7cm]{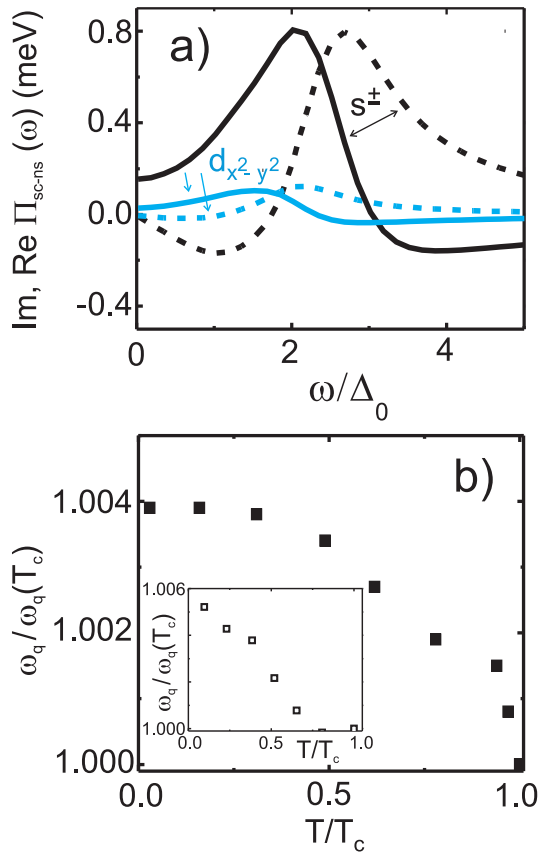}
\par
\caption{(color online) (a) Calculated difference of the polarization operator for the two $\alpha$-bands, $\Pi ({\bf q} \to 0, \omega)$  between the superconducting and normal states. The solid and dashed curves refer to the real and imaginary components, respectively. (b) Calculated normalized temperature dependent $B_{1g}$ frequencies for temperatures below T$_c$. The inset shows the experimental data for SKFA extracted from Fig.\ref{fig:4}.} \label{figth_1}
\end{figure}

The various symmetries of the superconducting gaps will differently renormalize the polarization operator in the superconducting state. The main effect of superconductivity on the phonon self-energy at {\bf q = 0} results from the change of the polarization operator of the two $\alpha$-bands. The latter which are centered close to the $\Gamma-$point and consequently couple rather strongly to the phonon dispersions around the center of the BZ. In Fig.\ref{figth_1}(a) we show the changes of the real and imaginary parts of the polarization operators of the $\alpha$-bands for the extended $s$-wave ($s^+$), and $d_{x^2-y^2}$-wave symmetries of the superconducting gap with respect to the normal state values.
One can clearly see that the largest effect occurs for the extended $s$-wave symmetry. The real part of the polarization operator in the superconducting state is larger for energies below $2\Delta_0$ than their normal state values and smaller for energies above $2\Delta_0$. This leads to a softening of the phonon frequencies with the corresponding sharpening of the phonon spectral function for those optical phonons which energies are lower than $2\Delta_0$. The opposite behavior is found for $\omega_{\Gamma}>2\Delta_0)$. This overall agrees well with the earlier Ref. \cite{Zeyher}. An interesting remark here is that although similar behavior is found for $d_{xy}$ (not shown) and $d_{x^2-y^2}$-wave symmetries it is much less pronounced due to the effectively smaller sizes of the gap at the Fermi surface of the $\alpha$-bands and the presence of the nodes. As a result the renormalization should be strongest for an extended $s$-wave symmetry. It is important to notice that the behavior of the polarization operator is not affected by the inclusion of the higher harmonics in the extended $s$-wave symmetry. The latter mostly affect the behavior of the gap around the $M$ points and not at the $\Gamma-$point where the coupling to the Raman-active phonons is present.

We also perform an analysis of the temperature dependence of the B$_{1g}$ mode for temperatures below the superconducting transition temperature assuming an extended $s^{\pm}$-wave symmetry. The respective results are shown in Fig.\ref{figth_1}(b). Using the previous estimates for the electron-phonon coupling strength of $g_{\bf q=0}=24.8$~meV and the bare $\hbar \omega_{B_{1g}} \approx 26.2$~meV in iron pnictides \cite{cohen}, we calculate the renormalization of the phonon frequency and find its hardening in the normal state at T$=100$~K to 26.3~meV which corresponds to 210.4~cm$^{-1}$. Below T$_c$ the Re$\Pi(\omega)$ further decreases at energies larger than $2\Delta_0 \approx 13.5$~meV which agrees well with the experimental data shown in the inset of Fig.\ref{figth_1}(b). At the same time, we also find a broadening of the phonon mode below T$_c$ (not shown) though experimentally our data show a quick switch of the broadening tendency into a sizable narrowing.
Such change may occur only if $\hbar \omega_{0} \sim 2\Delta_0$ or slightly smaller which
would yield to anomalously large $2\Delta_0/k_B T_c$ ratios. One of the possible explanations could be a microscopic coexistence of SDW and superconductivity or interband scattering effects which we leave for further studies. Evidence for coexistence, phase competition or separation on different length scales has indeed been claimed \cite{coexistence}. These effects seem to depend in a critical way on stoichiometry and therefore deserved further investigations.

Now we estimate the electron-phonon coupling strength. The linewidth of the isostructural compound FeAs$_2$ is given by $3-4~\mbox{cm}^{-1}$  between 4 K and room temperature. Taking into account the linewith of $3-7~\mbox{cm}^{-1}$  and the superconductivity related narrowing of $1-2~\mbox{cm}^{-1}$ in BKFA and SKFA, the electron-phonon contribution does not exceed $2~\mbox{cm}^{-1}$. The Allen equation provides a relation between the phonon linewidth, $\Gamma$ due to electron-phonon coupling and the phonon coupling constant~\cite{Allen}: $\Gamma=2\pi\lambda_{B_{1g}}N(0)\omega^2$ where $\lambda_{B_{1g}}$ is the strength of the electron-$B_{1g}$ coupling and $N(0)$ is the density of states (DOS) on the Fermi surface. In the BaFe$_2$As$_2$, the total DOS at $E_F$ is taken as $N(0)=3.06$ eV$^{-1}$/f.u.~\cite{Singh}. If we assume $N(0)$ remains constant for a small doping, the electron-phonon constant is estimated to be $\lambda_{B_{1g}}\approx 0.02$.

We can also obtain the electron-phonon constant at the Brillouin zone center using the superconductivity-induced renormalization constant $\kappa=(\omega^{SC}/\omega^{N})-1\approx 0.5\%$~\cite{Zeyher}: $\lambda_{B_{1g}}^{\Gamma-point}=-\kappa\mbox{Re}[(\sin u)/u]\approx 0.01$, where $u\equiv\pi +2i\cosh^{-1}(\omega^{N}/2\Delta)$. Both values are much smaller than the theoretically estimated total average value of $\lambda \approx 0.2$~\cite{Boeri} and our results are in agreement with weak electron-phonon coupling for the $B_{1g}$ phonon mode.

To conclude, we have presented a Raman scattering study of the undoped SrFe$_{2}$As$_{2}$  and superconducting BKFA and SKFA as a function of temperature. We observe a superconductivity induced self-energy effect of the B$_{1g}$ phonon
mode and estimate the corresponding electron-phonon strength of $\lambda_{B_{1g}}\approx 0.02$. In addition,
we find that an electronic continuum in the superconducting phase behaves like a nodeless gap function.

We acknowledge important discussions with B. Keimer and D. Wulferding. This work was supported by the German Science Foundation DFG and the ESF program \emph{Highly Frustrated Magnetism}. Work at the EPFL was supported by the Swiss NSF and by the NCCR MaNEP. KYC acknowledges financial support from the Alexander-von-Humboldt Foundation and the Korea Research Foundation Grant funded by the Korean Government (KRF-2008-313-C00247).


\end{document}